\documentclass[journal=jpclcd,manuscript=letter,layout=onecolumn]{achemso}
\pdfoutput=1
\usepackage{pdfpages}
\usepackage[version=3]{mhchem} 		
\usepackage{natmove}
\usepackage{bm}				
\newcommand{\Vector}[1]{\bm{#1}}	
\newcommand{\rv}{\Vector{r}}
\newcommand{\im}{\mathrm{i}}
\author{Caterina Cocchi}
\affiliation{Centro S3, CNR-Istituto Nanoscienze, I-41125 Modena, Italy}
\alsoaffiliation{Dipartimento di Fisica, Universit\`a di Modena e Reggio Emilia, I-41125 Modena, Italy}
\email{caterina.cocchi@unimore.it}
\author{Deborah Prezzi}
\affiliation{Centro S3, CNR-Istituto Nanoscienze, I-41125 Modena, Italy}
\email{deborah.prezzi@unimore.it}
\author{Alice Ruini}
\affiliation{Centro S3, CNR-Istituto Nanoscienze, I-41125 Modena, Italy}
\alsoaffiliation{Dipartimento di Fisica, Universit\`a di Modena e Reggio Emilia, I-41125 Modena, Italy}
\author{Enrico Benassi}
\affiliation{Centro S3, CNR-Istituto Nanoscienze, I-41125 Modena, Italy}
\author{Marilia J. Caldas}
\affiliation{Instituto de F{\'\i}sica, Universidade de S\~ao Paulo, 05508-900 S\~ao Paulo, SP, Brazil}
\author{Stefano Corni}
\affiliation{Centro S3, CNR-Istituto Nanoscienze, I-41125 Modena, Italy}
\author{Elisa Molinari}
\affiliation{Centro S3, CNR-Istituto Nanoscienze, I-41125 Modena, Italy}
\alsoaffiliation{Dipartimento di Fisica, Universit\`a di Modena e Reggio Emilia, I-41125 Modena, Italy}
\title{Optical Excitations and Field Enhancement in Short Graphene Nanoribbons}
\begin{document}
\newpage
\begin{abstract}
The optical excitations of elongated graphene nanoflakes of finite length are investigated theoretically through quantum chemistry semi-empirical approaches.
The spectra and the resulting dipole fields are analyzed, accounting in full atomistic details for quantum confinement effects, which are crucial in the nanoscale regime.
We find that the optical spectra of these nanostructures are dominated at low energy by excitations with strong intensity, comprised of characteristic coherent combinations of a few single-particle transitions with comparable weight. They give rise to stationary collective oscillations of the photoexcited carrier density extending throughout the flake, and to a strong dipole and field enhancement. This behavior is robust with respect to  width and length variations, thus ensuring tunability in a large frequency range. The implications for nanoantennas and other nanoplasmonic applications are discussed for realistic geometries.
\end{abstract}
\section*{TOC Graphic}
\begin{figure}%
\centering
\includegraphics[width=5.1cm]{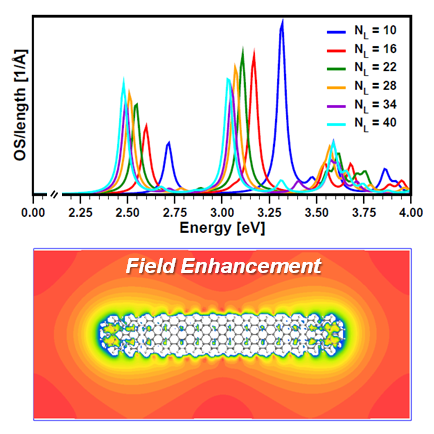}
\end{figure}

\textbf{Keywords}: nanoplasmonics, ZINDO, UV-vis spectrum, carbon nanostructures, transition density
\newpage
In the last few years remarkable interest has grown for nanoplasmonics and the perspectives it offers to merge electronics and photonics at the nanoscale \cite{ozba06sci,bron10sci}.
A wide range of potential applications can be designed, including sensing and spectroscopic techniques \cite{will-duyn07arpc,homo08cr,stew+08cr,anke+08natm}, fabrication of nanoantennas \cite{mueh+05sci,schn+09natpho,bhar+09aop,gao+11acr,gian+11cr} and light emitters \cite{berg-stoc03prl,stoc08natpho}, as well as beyond-THz optical devices or solar-energy conversion systems \cite{polm08sci,drag-drag08pqe,schu+10natm}.
While great attention has been devoted to metal nanoparticles, due to the relative ease to produce them and to their possibility to support surface modes \cite{kell+03jpcb,barn+03nat,lal+07natpho}, new materials and metamaterials are now being explored \cite{kons-sarg10natn,west+10lpr}, which, in addition to enhanced optical responses, are able to optimize circuit integration and reduce losses.
Graphene has proved to have unique electronic and mechanical properties \cite{cast+09rmp}, and has been more recently investigated also for photonics and optoelectronics \cite{beni09prb,avou10nl,bona+10natpho}. 
In the field of (nano)plasmonics, so far most emphasis has been devoted to spectroscopy of plasmons in extended graphene, either doped or undoped 
\cite{hwan-sarm07prb,erbe+08prb,kram+08prl,gass+08natn,jabl+09prb,mish+10prl,kopp+11nl}, and to the large lifetimes of their excitations compared to conventional metals \cite{jabl+09prb,thon+12nano}.
Some interesting predictions, mostly based on macroscopic models, have been proposed for plasmonics in spatially-modulated graphene of micron and sub-micron size range \cite{kopp+11nl,niki+11prb,vaki-engh11sci,wang+11prb,ju+11natn,chri+12nano}. 

\begin{figure}%
\centering
\includegraphics[width=.45\textwidth,clip]{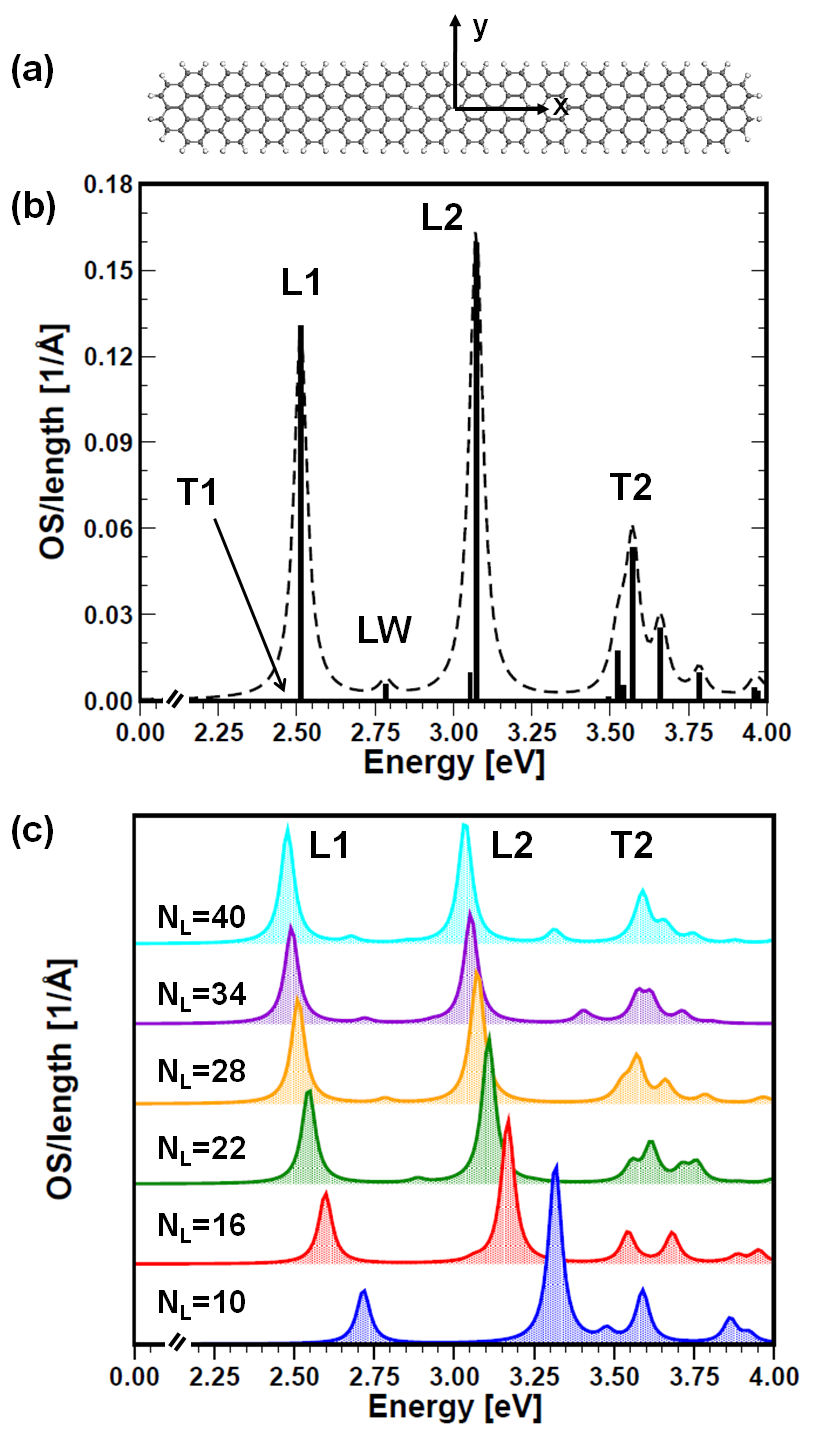}%
\caption{(a) Ball-and-stick model of a prototypical elongated graphene flake, characterized by length parameter $N_L=28$ ($x$ direction) and width parameter $N_W=7$ ($y$ direction). Edge atoms are passivated with H (white balls). (b) Optical spectrum (dashed line) calculated for the system in (a): the oscillator strength (OS) of individual excitations is indicated by black bars. The main excitations are labeled according to their polarization with respect to the long axis of the system ($x$), i.e. L for longitudinal and T for transverse polarization. 
(c) Optical spectra of a series of graphene flakes of fixed width ($N_W=7$) and variable length, from 24 to 88 \AA{}, labeled after their length parameter $N_L$, ranging from 10 to 40. The OS is normalized with respect to the flake length in both (b) and (c).
All spectra are obtained by introducing a Lorentzian broadening of 25 meV.
}
\label{fig1}
\end{figure}

Here we focus on graphene in a completely different regime, typical of nanoscale structures, where an optical gap opens as a consequence of quantum confinement, and we show that field enhancement effects can also be seen.
This regime has become particularly exciting in view of the recent production of controlled graphene wires by chemical routes \cite{wu+07cr,yang+08jacs,cai+10nat,palm-samo11natc}. 
We consider the case of self-standing elongated graphene nanoflakes (GNFs) with H-terminated edges, which we analyze by applying a fully-microscopic quantum-chemical approach.
These flakes can be thought of as finite portions of armchair-edged graphene nanoribbons (GNRs)\cite{naka+96prb}  [see \ref{fig1}(a)] of variable widths and lengths.
We focus on the optical properties, which are sensitive to the structural details of the system, namely length and width modulation, by computing their UV-vis spectra.
In the low energy region intense and tunable peaks are recognized, characterized by a coherent coupling of transitions that yield ``collective'' excitations with strong transition dipole.
The local field enhancement produced by these large dipole excitations is discussed for a prototypical flake of sub-nanometer width and nanometer length.

To investigate the nature of these optical excitations, their microscopic origin and their size-dependent trends, we adopt the semi-empirical Hartree-Fock-based method ZINDO, which implements the configuration-interaction (CI) procedure including single excitations only (S) \cite{ridl-zern73tca}.
This method is known to provide reliable results for the optical spectra of aromatic molecules \cite{cald+01apl,wang+04jacs,cocc+11jpcl} \bibnote{The reliability of the ZINDO model for the target systems was further demonstrated by computing time-dependent density-functional theory (TDDFT-b3lyp) excitation energies and oscillator strengths for a selected flake of length $\sim$ 24 \AA{} (GAMESS package with 3-21G basis set: \url{http://www.msg.ameslab.gov/gamess/})}.
All calculations are performed starting from optimized geometries obtained with AM1 \cite{dewa+85jacs,cocc+11jpcc}, used also for the calculation of mean-field ground state properties.

The analysis of the optical properties is carried out for a series of prototypical short graphene nanoribbons or \textit{graphene nanoflakes (GNFs)} of fixed width ($\sim$ 7.3 \AA{}) and variable length (from 24 to 88 \AA{}).
Following the standard notation for armchair GNRs \cite{naka+96prb}, these structures can be characterized by a width parameter $N_W$, which indicates the number of dimeric lines along the zigzag direction ($y$ axis). 
In addition, we here introduce a length parameter $N_L$ that corresponds to the number of zigzag chains in the armchair direction ($x$ axis), excluding the ends.
As shown in the model structure of \ref{fig1}(a), the flake edges are passivated with H atoms and the flake ends are shaped to minimize the zigzag-edge contributions \cite{shem+07apl,hod+08prb}.
The structure of a prototypical flake with $N_W = 7$ and $N_L = 28$ is displayed in \ref{fig1}(a) and its UV-vis spectrum (dashed line) is shown in \ref{fig1}(b), where the oscillator strengths (OS) of the individual optical excitations are indicated by black bars.
The optical spectrum is dominated by three intense peaks: the first two, L1 and L2, correspond to excitations with a large transition dipole along the $x$ axis of the system (L stands for "longitudinally polarized''); the third peak is found at higher energy and is dominated by the excitation T2, which shows a transverse polarization ($y$ direction) with respect to the long axis of the system.
In addition to these, we also find a transversally polarized excitation T1 with negligible oscillator strength below the first peak L1, and a weak longitudinally polarized excitation (LW) between L1 and L2.
\ref{fig1}(c) displays the calculated UV-vis spectra for GNFs of the same width ($N_W=7$) and increasing length, with $N_L$ ranging from 10 to 40. 
The main features described for the case of $N_L = 28$ are maintained along the flake series, except for L1 becoming the lowest-energy excitation in place of T1 for the longest GNF ($N_L=40$): this is a signature of the approaching behavior of quasi-1D armchair graphene nanoribbons (AGNRs) \cite{prez+08prb}.
Moreover, as a result of confinement, we find an overall red-shift of L1 and L2, with the optical gap decreasing from 3.14 to 2.48 eV. 
The energy \textit{difference} between these longitudinal excitations is almost unaffected upon flake length increase, ranging from about 0.55 eV ($N_L=40$) to 0.60 eV ($N_L=10$).
As expected, the energy of T1 and T2 does not vary considerably with length, since they are polarized along the flake width, kept fixed.
At increasing length, we also notice a significant transfer of OS to the lowest energy peak L1, which tends to saturate for the longest flake, again similarly to the behavior of infinite nanoribbons \cite{prez+08prb}.

\begin{figure}%
\centering
\includegraphics[width=.45\textwidth]{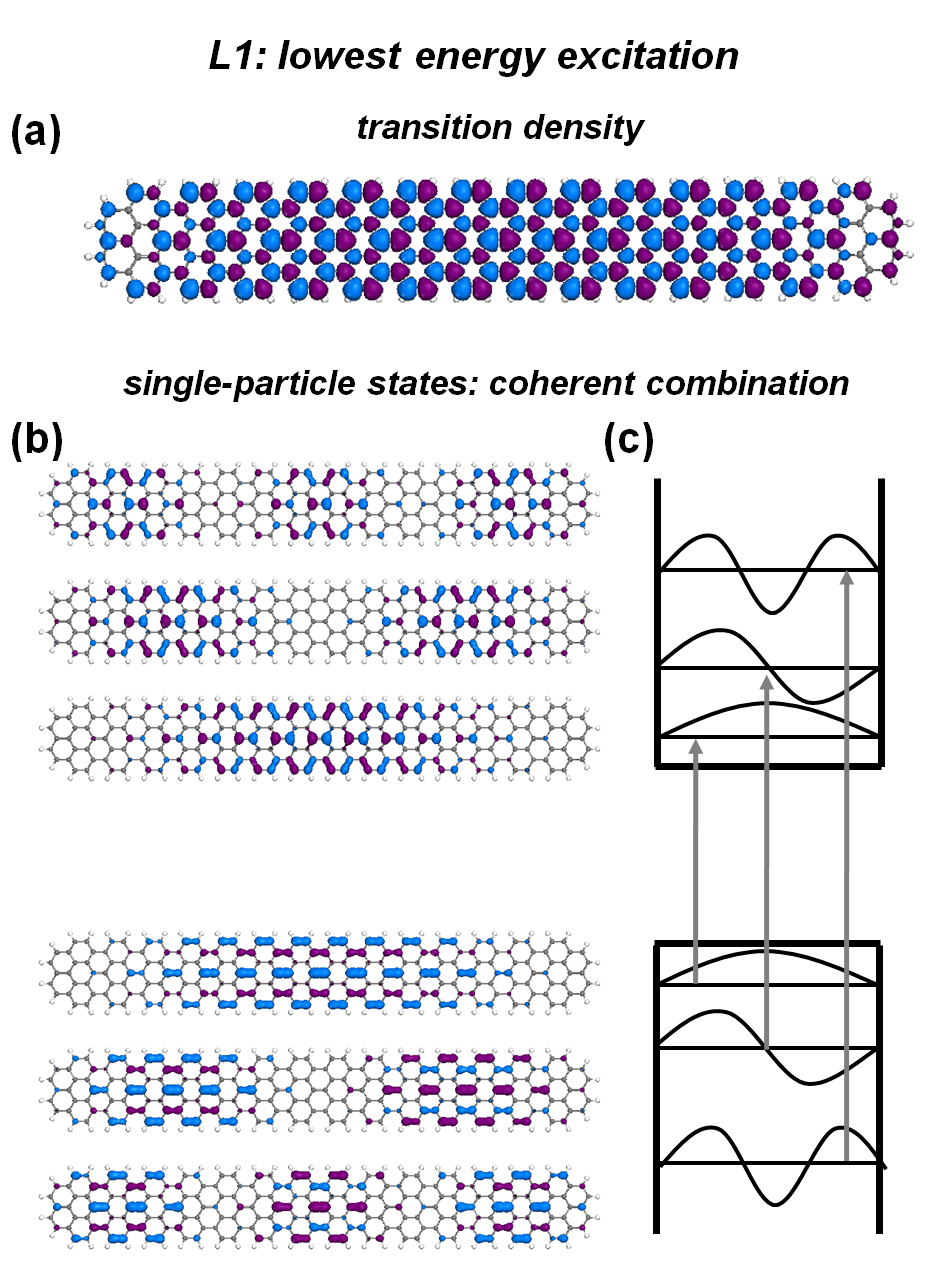}
\caption{Lowest energy excitation L1 of $N_L=28$ graphene flake shown in \ref{fig1}(a).
(a) Transition density computed according to \ref{td}. (b) Single-particle wave functions mostly contributing to L1.
The lowest energy excitation L1 is obtained from a coherent superposition of single particle transitions, as indicated in scheme (c).}
\label{fig2}
\end{figure}

\begin{figure}%
\centering
\includegraphics[width=.45\textwidth]{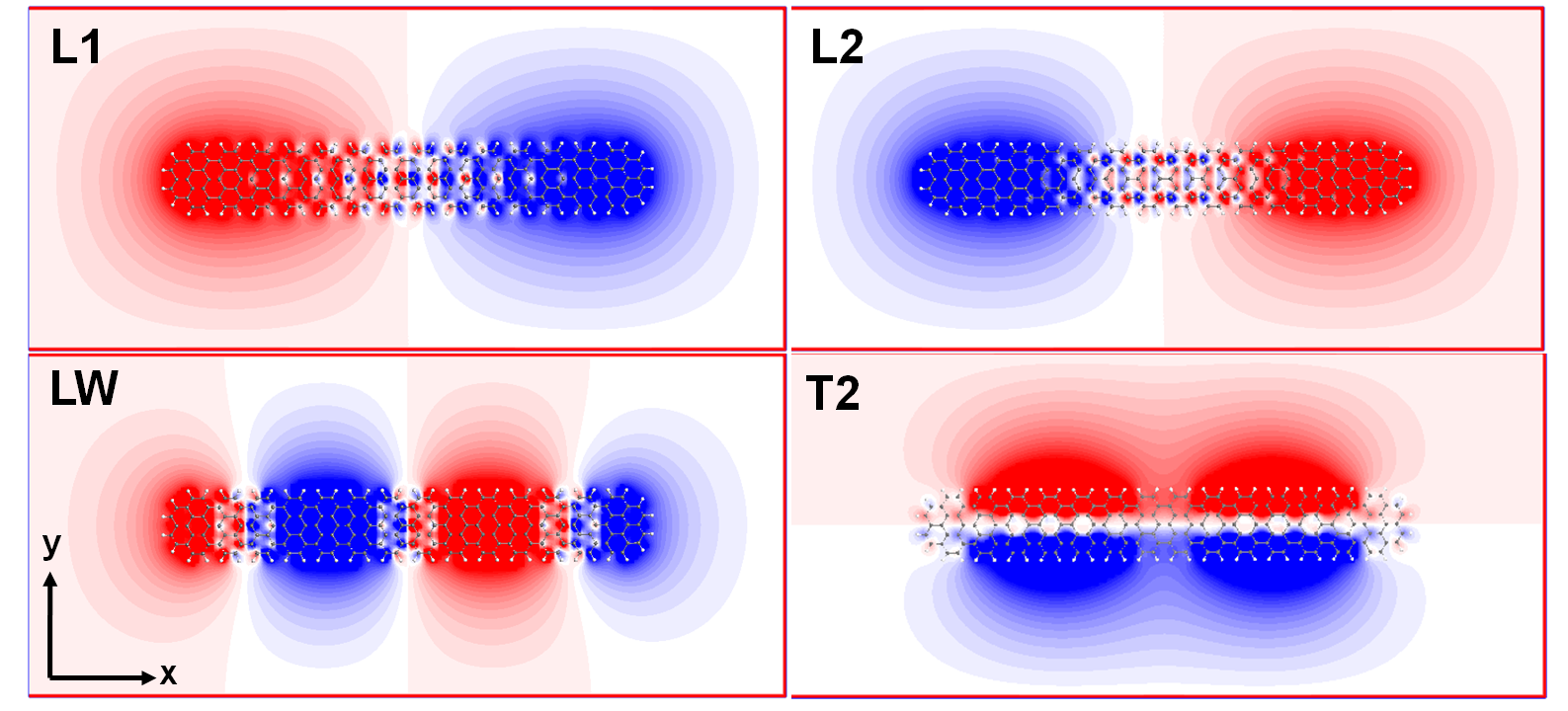}%
\caption{Pictorial view of the electric potential generated by the transition density relative to the excitations L1, L2, LW and T2 of the $N_L=28$ flake. }
\label{fig3}
\end{figure}

To gain further insight into the nature of these spectral features, we analyze the composition of all low-energy excitations in terms of molecular orbital (MO) transitions and CI weights. 
The numerical details are presented in the Supporting Information (Table S1) for few selected cases; here we focus again on $N_L = 28$, starting from the first bright excitation L1.
As illustrated in \ref{fig2}, this excitation arises from the coherent combination of single-particle transitions involving several harmonics of the same states.
Note that, while in shorter flakes the HOMO $\rightarrow$ LUMO transition contributes for most of the weight, in longer structures several higher harmonics enter the composition with comparable weights.
The fact that transitions between occupied and virtual MOs with the same envelope function modulation concur to form the excitation helps explaining the large OS observed, as this mechanism tends to maximize the wave function overlap. 
The spatial extension of the MOs increases with their energy distance from the frontier orbitals, further contributing to the large transition dipole. 
This effect is made particularly evident by plotting the transition density of the excitation, defined as:
\begin{equation}
\rho^{0p} (\rv) = \sum_{i a } C_{i a}^p \phi_{i}(\rv) \phi_{a}^*(\rv) ,
\label{td}
\end{equation}
where $C_{i a}^{p}$ are the CI coefficients of the $p$-th excited state, corresponding to single excitations from the occupied $\phi_{i}$ to virtual $\phi_{a}$ orbitals \bibnote{The approximations included within the adopted ZINDO/S approach, specifically in the neglect of two center integrals, have to be taken into account in the calculation of $\rho^{0p} (\rv)$.}.
From \ref{fig2}(a) we notice that the density extends homogeneously over the entire structure with an underlying dipolar character, and is not concentrated just to the central portion.
This is evident also by inspecting the contour plot of the corresponding electric potential, pictorially represented in \ref{fig3}.

Consistently with their polarization, L1 and L2 have a large dipole oriented along $x$, while that of T2 is along $y$. 
As shown for L1 in \ref{fig2}, also L2 and T2 arise from combinations of transitions between states with the same envelope function modulation.
On the contrary, LW is mainly composed by transitions involving MOs with different number of nodes. 
This gives rise to a multipolar modulation in the electric potential, as shown in \ref{fig3}, and leads to the weaker intensity observed for this excitation.

\begin{figure}%
\centering
\includegraphics[width=.45\textwidth]{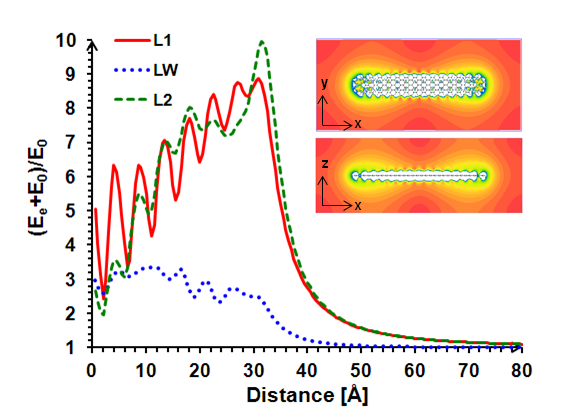}%
\caption{Field enhancement of longitudinally polarized excitations L1, L2 and LW of $N_W$=7 and $N_L$=28 graphene nanoflake computed along the flake axis, at 3.3 \AA{} from its plane. Close to the flake center, above its surface, the field enhancement is susceptible of the transition density modulation [see \ref{fig2}(a)]. In the inset, pictorial sketch of the field enhancement contour plot in ($x$,$y$) and ($x$,$z$) views.
Notice that the values of the field enhancement are sensitive to the choice of $\Gamma_p$, according to \ref{eq:rhoapp}: here a conservative value $\Gamma_p$=25 meV is adopted.
}
\label{fig4}
\end{figure}

\begin{figure}%
\centering
\includegraphics[width=.45\textwidth]{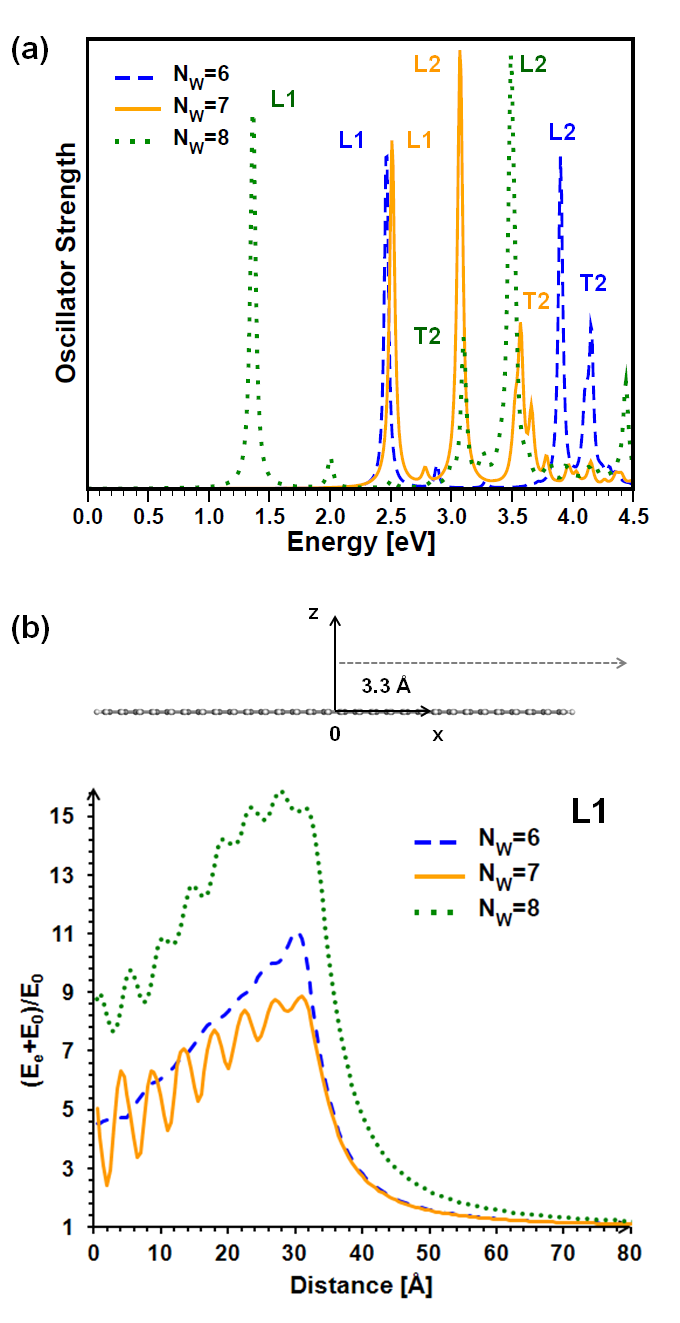}%
\caption{(a) UV-vis spectra of a series of graphene flake of fixed length ($N_L=28$) and variable width, with $N_W$ ranging from 6 to 8. A Lorentzian broadening of 25 meV is introduced.
(b) Field enhancement of L1 excitations along the flake longitudinal axis, at a distance of 3.3 \AA{} from its basal plane, according to the scheme reported on top of the graph.
Notice that the values of the field enhancement are sensitive to the choice of $\Gamma_p$, according to \ref{eq:rhoapp}: here a conservative value $\Gamma_p$=25 meV is adopted.
}
\label{fig5}
\end{figure}

The large dipole strength and the ``beyond-single-particle'' nature of L1 and L2 recall the main features related to collective plasmonic excitations 
in small metal clusters, as analyzed by means of atomistic methods \cite{yan+07prl,yan-gao08prb,lian+09jcp,yasu+11pra}.
We thus calculate for our systems a quantity which is usually adopted for the characterization of plasmonic excitations, i.e. the local enhancement of an electromagnetic field incident on the system.
To this end, we approximate the flake response function, at resonance conditions, via the electronic density variation $\delta \rho (\omega_{0p}; \rv)$ induced by the periodic external electric field $\Vector{E}_{0} exp(-\im\omega_{0p}t)$, given by:
\begin{equation}
\delta \rho (\omega_{0p}; \rv)= -\frac{\rho^{0p}(\rv)}{\im\hbar\Gamma_p}
 \Vector{\mu}^{0p} \cdot \Vector{E}_{0} .
\label{eq:rhoapp}
\end{equation}
Here $\Vector{\mu}^{0p}$ is the transition dipole for the excitation to the in-resonance $p$ state, and $\Gamma_p$ is the corresponding decay rate, related to the intrinsic absorption bandwidth. 
Details on the derivation of \ref{eq:rhoapp} are given in the Supporting Information. 
The oscillating charge distribution $\delta \rho (\omega_{0p}; \rv)$ originates an additional electric field $\Vector{E}_e (\rv) exp(-\im\omega_{0p}t)$ leading to an overall enhanced electric field ($E_0+E_e$) nearby the graphene flake. 
The transition dipole $\Vector{\mu}^{0p}$ is a straight result from the calculation, while $\Gamma_p$ is a critical parameter in determining the maximum enhancement [the larger this value, the smaller the enhancement, as shown in \ref{eq:rhoapp}]. 
The field enhancement is obtained by deriving the electric Coulomb potential generated by the transition density, and expressed in units of $E_0$ ($E_e/E_0$).
Since experimental data are not yet available for graphene nanoflakes such as those examined here, we adopt in the figures a conservative value $\Gamma_p$=25 meV, chosen from the spectral linewidths measured for single-wall carbon nanotubes (SWCNTs) at room temperature in solution \cite{scho06natm}.
Due to the critical dependence of the field enhancement on the choice of $\Gamma_p$ [see \ref{eq:rhoapp}], it is worth noting that the field enhancement shown in \ref{fig4} would increase of over one order of magnitude for $\Gamma_p \sim$ 1 meV, as found for suspended SWCNTs at low temperature \cite{mats+08prb}.
In \ref{fig4} we show the field enhancement produced by the three main longitudinal excitations (L1, LW and L2) identified in \ref{fig1}(b) for the GNF of $N_W$=7 and $N_L$=28, computed along the flake longitudinal ($x$) axis, at a distance of 3.3 \AA{} from the basal plane [see \ref{fig5}(b)].
The oscillating character of the curves in the region above the flake surface is due to the transition density modulation [see \ref{fig2}(a)].
Beyond the flake border the oscillations disappear and, at distance much larger than the flake half-length ($\sim$30 \AA{}), the field enhancement assumes a Coulomb-like decay, proportional to $1/r^2$.
According to the OS of the corresponding excitations, the field enhancement for L1 and L2 have comparable values, while that of LW is about three times lower.

Finally, we discuss the stability and the tunability of these optical properties with respect to length and width modulation. 
Focusing specifically on the lowest energy peak L1, the computed values of field enhancement are basically independent of the ribbon length (see Supporting Information, Figure S1).
On the other hand it is well known that both the electronic and optical properties of quasi-1D AGNRs are sensitive to width modulation \cite{son+06prl,baro+06nl,yang+07nl,prez+08prb,prez+11prb}.
Also for the finite flakes considered here three families are identified \cite{cocc+11jpcc}, characterized by different electronic gaps, with the smallest values pertaining to the $N_W = 3p+2$ family ($p$ integer).
Here we investigate two additional graphene flakes of width parameters $N_W$=6 and $N_W$=8, keeping their length fixed at $N_L$=28.
As shown in \ref{fig5}(a), the three main peaks (L1, L2 and T2) observed for $N_W$=7 are preserved in all families.
While T2 redshifts at increasing width, as expected, the energy and the intensity of the longitudinal excitations L1 and L2 are closely related to the electronic properties of each family.
In particular the trend for the electronic gaps \cite{cocc+11jpcc} is reflected in the excitation energies of L1.
The field enhancement computed for the L1 excitation of each GNF of different width is shown in \ref{fig5}(b): we find a distinctive enhancement for $N_W$=8 compared to the other two, which already comes from the transition dipole moments of the individual single-particle excitations.

In summary, we have analyzed the optical excitations of finite graphene nanoribbons, of sub-nanometer width and nanometer length.
At low energy the UV-vis spectra are dominated by intense excitations with longitudinal polarization with respect to the flake long axis.
These are characterized by a ``collective'' character, coming from coherent superposition of MO transitions with the same envelope function modulation.
The investigated excitations are tunable both in energy and intensity upon appropriate length and width ribbon modulation.
The field enhancement computed for these excitations suggests the applicability of these systems as nanoantennas and in other optoelectronic and nanoplasmonic applications.

\begin{acknowledgement}
The authors are grateful to Andrea Bertoni, Ulrich Hohenester and Massimo Rontani for stimulating and helpful discussions, and acknowledge CINECA for computational support. 
This work was partly supported by the Italian Ministry of University and Research under FIRB grant ItalNanoNet, and by Fondazione Cassa
di Risparmio di Modena with project COLDandFEW.
M.~J.~C.~ acknowledges support from FAPESP and CNPq (Brazil).
\end{acknowledgement}
\begin{suppinfo}
We include the description of the main excitations, both longitudinally and transversally polarized, for selected graphene ribbons of variable length and width.
The effects of length modulation on the field enhancement produced by the lowest energy bright excitation are discussed, supported by a figure showing the corresponding curves for few selected cases.
Finally, the analytical expression of the field enhancement is analytically derived from the transition density on the basis of linear response theory.
\end{suppinfo}
\providecommand*{\mcitethebibliography}{\thebibliography}
\csname @ifundefined\endcsname{endmcitethebibliography}
{\let\endmcitethebibliography\endthebibliography}{}

  \clearpage
  

\section*{Supplementary material}

# SUPPORTING INFORMATION

# Optical Excitations and Field Enhancement in Short Graphene Nanoribbons

Caterina Cocchi,*,†,‡ Deborah Prezzi,*,† Alice Ruini,†,‡ Enrico Benassi,† Marilia J. Caldas,¶ Stefano Corni,† and Elisa Molinari†,‡

*Centro S3, CNR-Istituto Nanoscienze, I-41125 Modena, Italy, Dipartimento di Fisica, Università di Modena e Reggio Emilia, I-41125 Modena, Italy, and Instituto de Física, Universidade de São Paulo, 05508-900 São Paulo, SP, Brazil*

E-mail: caterina.cocchi@unimore.it; deborah.prezzi@unimore.it

---
*To whom correspondence should be addressed
†Centro S3, CNR-Istituto Nanoscienze, I-41125 Modena, Italy
‡Dipartimento di Fisica, Università di Modena e Reggio Emilia, I-41125 Modena, Italy
¶Instituto de Física, Universidade de São Paulo, 05508-900 São Paulo, SP, Brazil

# Excitation Analysis in Terms of Orbital Transitions and Weights

In Table S1 we report the lowest energy excitations, labeled after Figure 1(b) in the main text, of three graphene nanoflakes (GNFs) at increasing length. We consider the shortest ($N_L$=10, $\sim$ 24 Å) and the longest ($N_L$=40, $\sim$ 88 Å) ones, in addition to $N_L$=28 GNF (length $\sim$ 63 Å) discussed in details in the main text. The excitation energy, oscillator strength (OS) and composition in terms of molecular orbital (MO) transitions for the main excitations are presented.

From the results presented in Table S1, the gain of OS characterizing the first bright excitation L1 at increasing length is evident. It is also worth noting that for the longest considered ribbon ($N_L$=40) the lowest energy excitation is indeed L1, being basically degenerate with T1.

| System | Excitation | Energy [eV] | OS | Transitions (weight) |
|---|---|---|---|---|
| $N_L$=10 | T1 | 2.54 | 0.0005 | HOMO-1 $\to$ LUMO (0.37) |
| | | | | HOMO $\to$ LUMO+1 (0.39) |
| | L1 | 2.72 | 1.62 | HOMO $\to$ LUMO (0.79) |
| | L2 | 3.32 | 5.43 | HOMO-3 $\to$ LUMO+3 (0.12) |
| | | | | HOMO-1 $\to$ LUMO+1 (0.76) |
| | LW | 3.48 | 0.34 | HOMO-4 $\to$ LUMO (0.14) |
| | | | | HOMO-2 $\to$ LUMO+2 (0.26) |
| | | | | HOMO $\to$ LUMO+4 (0.31) |
| | T2 | 3.59 | 1.56 | HOMO-1 $\to$ LUMO (0.41) |
| | | | | HOMO $\to$ LUMO+1 (0.39) |

*Continued on next page*

| System | Excitation | Energy [eV] | OS | Transitions (weight) |
|---|---|---|---|---|
| $N_L$=14 | T1 | 2.48 | 0.00 | HOMO-3 → LUMO (0.21) |
| | | | | HOMO → LUMO+3 (0.22) |
| | L1 | 2.51 | 8.19 | HOMO-2 → LUMO+2 (0.11) |
| | | | | HOMO-1 → LUMO+1 (0.21) |
| | | | | HOMO → LUMO (0.48) |
| | LW | 2.78 | 0.35 | HOMO-2 → LUMO (0.21) |
| | | | | HOMO-1 → LUMO+1 (0.15) |
| | | | | HOMO → LUMO+2 (0.23) |
| | L2 | 3.07 | 10.00 | HOMO-6 → LUMO+6 (0.12) |
| | | | | HOMO-4 → LUMO+4 (0.20) |
| | | | | HOMO-3 → LUMO+3 (0.48) |
| | T2 | 3.57 | 3.35 | HOMO-4 → LUMO+2 (0.16) |
| | | | | HOMO-3 → LUMO (0.18) |
| | | | | HOMO-1 → LUMO+4 (0.15) |
| | | | | HOMO → LUMO+3 (0.17) |

| System | Excitation | Energy [eV] | OS | Transitions (weight) |
|---|---|---|---|---|
| $N_L$=40 | L1 | 2.48 | 12.56 | HOMO-2 → LUMO+2 (0.13) |
| | | | | HOMO-1 → LUMO+1 (0.19) |
| | | | | HOMO → LUMO (0.36) |
| | T1 | 2.48 | 0.00 | HOMO-4 → LUMO (0.15) |
| | | | | HOMO → LUMO+4 (0.16) |
| | LW | 2.68 | 0.62 | HOMO-2 → LUMO (0.18) |
| | | | | HOMO-1 → LUMO+1 (0.11) |
| | | | | HOMO → LUMO+3 (0.19) |
| | L2 | 3.03 | 13.30 | HOMO-7 → LUMO+7 (0.12) |
| | | | | HOMO-5 → LUMO+5 (0.19) |
| | | | | HOMO-4 → LUMO+4 (0.35) |
| | T2 | 3.59 | 5.16 | HOMO-4 → LUMO (0.14) |
| | | | | HOMO → LUMO+4 (0.13) |

Table S1: Energy, oscillator strength (OS) and composition of the main excitations characterizing the spectra of $N_W$=7 GNFs of length $N_L$=10, $N_L$=28 and $N_L$=40. In the last column we include the MO transitions with relative weights larger than 0.1.

In Table S2 we report the lowest energy excitations of $N_W$=6, $N_W$=7 and $N_W$=8 graphene flakes of length $N_L$=28. Excitation L1 in the largest flake ($N_W$=8) presents an increased OS of about 10% with respect to $N_W$=6 and $N_W$=7 GNF, where it keeps almost the same intensity. Also the energy of L1 is red shifted of over 1 eV in $N_W$=8 GNF, with respect to that of the other two, as observed in the UV-vis spectra of Figure 5(a) of the main text.

Table S2: Energy, oscillator strength (OS) and composition of the two lowest energy excitations of graphene nanoflakes with $N_L$=28 and different width, namely $N_W$=6, $N_W$=7 and $N_W$=8. In the last column only we include the MO transitions with relative weights larger than 0.1.

| System | Excitation | Energy [eV] | OS | Transitions (weight) |
|---|---|---|---|---|
| $N_W$=6 | L1 | 2.47 | 8.23 | HOMO-1 → LUMO+1 (0.20) <br> HOMO → LUMO (0.54) |
| $N_W$=7 | T1 | 2.48 | 0.002 | HOMO-3 → LUMO (0.21) <br> HOMO → LUMO+3 (0.22) |
| | L1 | 2.51 | 8.19 | HOMO-2 → LUMO+2 (0.11) <br> HOMO-1 → LUMO+1 (0.21) <br> HOMO → LUMO (0.48) |
| $N_W$=8 | L1 | 1.37 | 9.15 | HOMO-1 → LUMO+1 (0.18) <br> HOMO → LUMO (0.69) |

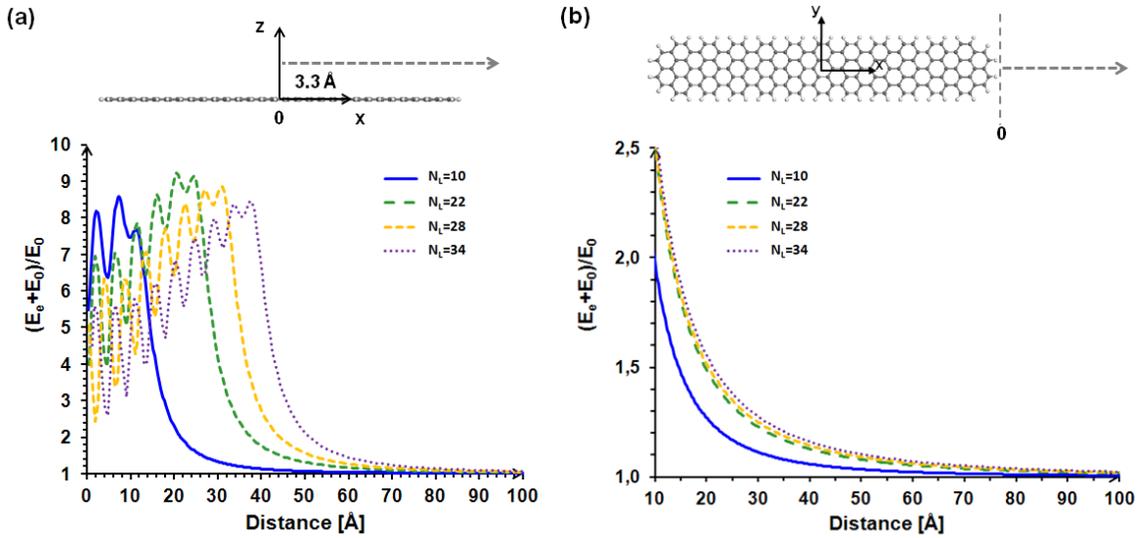

Figure S1: Field enhancement of longitudinally polarized excitation L1 for selected graphene nanoflakes of fixed width $N_W$=7 and increasing length from about 24 ($N_L$=10) to about 75 Å ($N_L$=34), computed along the flake axis (a) at 3.3 Å from its plane and (b) along the ribbon plane, starting from its lateral end. Notice that the values of the field enhancement are sensitive to the choice of $\Gamma_p$, according to Eq. (2) in the main text: here a conservative value $\Gamma_p$=25 meV is adopted.

## Effects of Length Increase on the Field Enhancement

We report in Figure S1 the field enhancement produced by L1 excitation along the longitudinal (*x*) axis of selected graphene flakes of increasing length, at different heights from the basal plane: at 3.3 Å above the plane from the flake center [see Figure S1(a)] and along the plane from the flake end [see Figure S1(b)]. It is worth noting that above the ribbon plane the field enhancement is basically independent of the length, and the curves are characterized by an oscillating behavior related to the transition density modulation. On the other hand, further from the flake lateral border, where the dipolar character of the transition density becomes predominant [see also Figure 2(a) in the main text], the field enhancement curves follow the oscillator strength gain of L1 upon increasing length [see Table S1 and also Figure 1(c) in the main text].

## Enhanced Response Field from Transition Density

We derive the relation between the enhanced field and the transition properties of the molecule. In general, the electric field $\vec{E}_e$ produced by the polarization of the molecule due to an external potential $V_0$ oscillating at a frequency $\omega$ (for example, due to an external electric field $\vec{E}_0$ considered in the quasi-static limit) can be found from:

$$\nabla \cdot \vec{E}_e = 4\pi \delta \rho(\vec{r}) \qquad (1)$$

where $\delta \rho(\vec{r})$ is the electronic density induced in the position $\vec{r}$ by $\vec{E}_0$. From response theory, it comes:[1]

$$\delta \rho(\omega; \vec{r}) = \int \chi(\omega; \vec{r}, \vec{r}') V_0(\vec{r}') \, d\vec{r}', \qquad (2)$$

where the molecular response function can be written as:

$$\chi(\omega; \vec{r}, \vec{r}') = \sum_{i \neq 0} \frac{\rho_{0i}(\vec{r}) \rho_{i0}(\vec{r}')}{\hbar\omega + i\hbar\Gamma_i + E_i - E_0} + \frac{\rho_{i0}(\vec{r}) \rho_{0i}(\vec{r}')}{-\hbar\omega - i\hbar\Gamma_i + E_i - E_0} \qquad (3)$$

In Eq. (3), $\rho_{0i}$ are the transition densities between the ground state 0 and the excited state $i$, and $E_i$ are the energies of the states. When the frequency $\omega$ is close to a molecular transition from 0 to $p$ (i.e., $\omega \approx \omega_{p0} = (E_p - E_0)/\hbar$), one denominator in Eq. (3) becomes very small, and the corresponding term dominates the sum over $i$, so that:

$$\chi(\omega_{p0}; \vec{r}, \vec{r}') \approx \frac{\rho_{0p}(\vec{r}) \rho_{p0}(\vec{r}')}{i\hbar \Gamma_p}. \quad (4)$$

Therefore:

$$\delta\rho(\omega; \vec{r}) = \frac{\rho_{0p}(\vec{r})}{i\hbar \Gamma_p} \int \rho_{p0}(\vec{r}') V_0(\vec{r}') \, d\vec{r}'. \quad (5)$$

If we now assume that $V_0(\vec{r}')$ is the electrostatic potential associated with an external probing field in the quasi static limit, then:

$$V_0(\vec{r}') = -\vec{r}' \cdot \vec{E}_0 \quad (6)$$

and

$$\delta\rho(\omega_{0p}; \vec{r}) = -\frac{\rho_{0p}(\vec{r})}{i\hbar \Gamma_p} \vec{\mu}_{p0} \cdot \vec{E}_0, \quad (7)$$

where $\vec{\mu}_{p0}$ is the molecular transition dipole, evaluated here by the ZINDO/S method. By solving Eq. (1) using the expression of $\delta\rho(\omega; \vec{r})$ in Eq. (7), the electric potential and the field $\vec{E}_e$ plotted in Figure 3 and Figure 4 of the main text are finally obtained.

## References

(1) Fetter, A. L.; Walecka, J. D. *Quantum theory of many-particle systems*; McGraw-Hill, New York, 1971.




\begin{mcitethebibliography}{65}
\providecommand*{\natexlab}[1]{#1}
\providecommand*{\mciteSetBstSublistMode}[1]{}
\providecommand*{\mciteSetBstMaxWidthForm}[2]{}
\providecommand*{\mciteBstWouldAddEndPuncttrue}
  {\def\EndOfBibitem{\unskip.}}
\providecommand*{\mciteBstWouldAddEndPunctfalse}
  {\let\EndOfBibitem\relax}
\providecommand*{\mciteSetBstMidEndSepPunct}[3]{}
\providecommand*{\mciteSetBstSublistLabelBeginEnd}[3]{}
\providecommand*{\EndOfBibitem}{}
\mciteSetBstSublistMode{f}
\mciteSetBstMaxWidthForm{subitem}{(\alph{mcitesubitemcount})}
\mciteSetBstSublistLabelBeginEnd{\mcitemaxwidthsubitemform\space}
{\relax}{\relax}

\bibitem[Ozbay(2006)]{ozba06sci}
Ozbay,~E. Plasmonics: Merging Photonics and Electronics at Nanoscale
  Dimensions. \emph{Science} \textbf{2006}, \emph{311}, 189--193\relax
\mciteBstWouldAddEndPuncttrue
\mciteSetBstMidEndSepPunct{\mcitedefaultmidpunct}
{\mcitedefaultendpunct}{\mcitedefaultseppunct}\relax
\EndOfBibitem
\bibitem[Brongersma and Shalaev(2010)]{bron10sci}
Brongersma,~M.~L.; Shalaev,~V.~M. The Case for Plasmonics. \emph{Science}
  \textbf{2010}, \emph{328}, 440--441\relax
\mciteBstWouldAddEndPuncttrue
\mciteSetBstMidEndSepPunct{\mcitedefaultmidpunct}
{\mcitedefaultendpunct}{\mcitedefaultseppunct}\relax
\EndOfBibitem
\bibitem[Willets and Van~Duyne(2007)]{will-duyn07arpc}
Willets,~K.; Van~Duyne,~R. Localized Surface Plasmon Resonance Spectroscopy and
  Sensing. \emph{Annu. Rev. Phys. Chem.} \textbf{2007}, \emph{58},
  267--297\relax
\mciteBstWouldAddEndPuncttrue
\mciteSetBstMidEndSepPunct{\mcitedefaultmidpunct}
{\mcitedefaultendpunct}{\mcitedefaultseppunct}\relax
\EndOfBibitem
\bibitem[Homola(2008)]{homo08cr}
Homola,~J. Surface Plasmon Resonance Sensors for Detection of Chemical and
  Biological Species. \emph{Chem.~Rev.~} \textbf{2008}, \emph{108},
  462--493\relax
\mciteBstWouldAddEndPuncttrue
\mciteSetBstMidEndSepPunct{\mcitedefaultmidpunct}
{\mcitedefaultendpunct}{\mcitedefaultseppunct}\relax
\EndOfBibitem
\bibitem[Stewart et~al.(2008)Stewart, Anderton, Thompson, Maria, Gray, Rogers,
  and Nuzzo]{stew+08cr}
Stewart,~M.~E.; Anderton,~C.~R.; Thompson,~L.~B.; Maria,~J.; Gray,~S.~K.;
  Rogers,~J.~A.; Nuzzo,~R.~G. Nanostructured Plasmonic Sensors.
  \emph{Chem.~Rev.~} \textbf{2008}, \emph{108}, 494--521\relax
\mciteBstWouldAddEndPuncttrue
\mciteSetBstMidEndSepPunct{\mcitedefaultmidpunct}
{\mcitedefaultendpunct}{\mcitedefaultseppunct}\relax
\EndOfBibitem
\bibitem[Anker et~al.(2008)Anker, Hall, Lyandres, Shah, Zhao, and
  Van~Duyne]{anke+08natm}
Anker,~J.; Hall,~W.; Lyandres,~O.; Shah,~N.; Zhao,~J.; Van~Duyne,~R. Biosensing
  with Plasmonic Nanosensors. \emph{Nature Mat.} \textbf{2008}, \emph{7},
  442--453\relax
\mciteBstWouldAddEndPuncttrue
\mciteSetBstMidEndSepPunct{\mcitedefaultmidpunct}
{\mcitedefaultendpunct}{\mcitedefaultseppunct}\relax
\EndOfBibitem
\bibitem[Muehlschlegel et~al.(2005)Muehlschlegel, Eisler, Martin, Hecht, and
  Pohl]{mueh+05sci}
Muehlschlegel,~P.; Eisler,~H.-J.; Martin,~O. J.~F.; Hecht,~B.; Pohl,~D.~W.
  Resonant Optical Antennas. \emph{Science} \textbf{2005}, \emph{308},
  1607--1609\relax
\mciteBstWouldAddEndPuncttrue
\mciteSetBstMidEndSepPunct{\mcitedefaultmidpunct}
{\mcitedefaultendpunct}{\mcitedefaultseppunct}\relax
\EndOfBibitem
\bibitem[Schnell et~al.(2009)Schnell, Garcia-Etxarri, Huber, Crozier, Aizpurua,
  and Hillenbrand]{schn+09natpho}
Schnell,~M.; Garcia-Etxarri,~A.; Huber,~A.; Crozier,~K.; Aizpurua,~J.;
  Hillenbrand,~R. Controlling the Near-Field Oscillations of Loaded Plasmonic
  nanoantennas. \emph{Nat. Photonics} \textbf{2009}, \emph{3}, 287--291\relax
\mciteBstWouldAddEndPuncttrue
\mciteSetBstMidEndSepPunct{\mcitedefaultmidpunct}
{\mcitedefaultendpunct}{\mcitedefaultseppunct}\relax
\EndOfBibitem
\bibitem[Bharadwaj et~al.(2009)Bharadwaj, Deutsch, and Novotny]{bhar+09aop}
Bharadwaj,~P.; Deutsch,~B.; Novotny,~L. Optical Antennas.
  \emph{Adv.~Opt.~Photon.~} \textbf{2009}, \emph{1}, 438--483\relax
\mciteBstWouldAddEndPuncttrue
\mciteSetBstMidEndSepPunct{\mcitedefaultmidpunct}
{\mcitedefaultendpunct}{\mcitedefaultseppunct}\relax
\EndOfBibitem
\bibitem[Gao et~al.(2011)Gao, Ueno, and Misawa]{gao+11acr}
Gao,~S.; Ueno,~K.; Misawa,~H. Plasmonic Antenna Effects on Photochemical
  Reactions. \emph{Acc.~Chem.~Res.~} \textbf{2011}, \emph{44}, 251--260\relax
\mciteBstWouldAddEndPuncttrue
\mciteSetBstMidEndSepPunct{\mcitedefaultmidpunct}
{\mcitedefaultendpunct}{\mcitedefaultseppunct}\relax
\EndOfBibitem
\bibitem[Giannini et~al.(2011)Giannini, Fernandez-Dominguez, Heck, and
  Maier]{gian+11cr}
Giannini,~V.; Fernandez-Dominguez,~A.~I.; Heck,~S.~C.; Maier,~S.~A. Plasmonic
  Nanoantennas: Fundamentals and Their Use in Controlling the Radiative
  Properties of Nanoemitters. \emph{Chem.~Rev.~} \textbf{2011}, \emph{111},
  3888--3912\relax
\mciteBstWouldAddEndPuncttrue
\mciteSetBstMidEndSepPunct{\mcitedefaultmidpunct}
{\mcitedefaultendpunct}{\mcitedefaultseppunct}\relax
\EndOfBibitem
\bibitem[Bergman and Stockman(2003)]{berg-stoc03prl}
Bergman,~D.~J.; Stockman,~M.~I. Surface Plasmon Amplification by Stimulated
  Emission of Radiation: Quantum Generation of Coherent Surface Plasmons in
  Nanosystems. \emph{Phys. Rev. Lett.} \textbf{2003}, \emph{90}, 027402\relax
\mciteBstWouldAddEndPuncttrue
\mciteSetBstMidEndSepPunct{\mcitedefaultmidpunct}
{\mcitedefaultendpunct}{\mcitedefaultseppunct}\relax
\EndOfBibitem
\bibitem[Stockman(2008)]{stoc08natpho}
Stockman,~M. Spasers Explained. \emph{Nat. Photonics} \textbf{2008}, \emph{2},
  327--329\relax
\mciteBstWouldAddEndPuncttrue
\mciteSetBstMidEndSepPunct{\mcitedefaultmidpunct}
{\mcitedefaultendpunct}{\mcitedefaultseppunct}\relax
\EndOfBibitem
\bibitem[Polman(2008)]{polm08sci}
Polman,~A. Plasmonics Applied. \emph{Science} \textbf{2008}, \emph{322},
  868--869\relax
\mciteBstWouldAddEndPuncttrue
\mciteSetBstMidEndSepPunct{\mcitedefaultmidpunct}
{\mcitedefaultendpunct}{\mcitedefaultseppunct}\relax
\EndOfBibitem
\bibitem[Dragoman and Dragoman(2008)]{drag-drag08pqe}
Dragoman,~M.; Dragoman,~D. Plasmonics: Applications to Nanoscale Terahertz and
  Optical Devices. \emph{Prog. Quantum Electron.} \textbf{2008}, \emph{32}, 1
  -- 41\relax
\mciteBstWouldAddEndPuncttrue
\mciteSetBstMidEndSepPunct{\mcitedefaultmidpunct}
{\mcitedefaultendpunct}{\mcitedefaultseppunct}\relax
\EndOfBibitem
\bibitem[Schuller et~al.(2010)Schuller, Barnard, Cai, Jun, White, and
  Brongersma]{schu+10natm}
Schuller,~J.; Barnard,~E.; Cai,~W.; Jun,~Y.; White,~J.; Brongersma,~M.
  Plasmonics for Extreme Light Concentration and Manipulation. \emph{Nature
  Mat.} \textbf{2010}, \emph{9}, 193--204\relax
\mciteBstWouldAddEndPuncttrue
\mciteSetBstMidEndSepPunct{\mcitedefaultmidpunct}
{\mcitedefaultendpunct}{\mcitedefaultseppunct}\relax
\EndOfBibitem
\bibitem[Kelly et~al.(2003)Kelly, Coronado, Zhao, and Schatz]{kell+03jpcb}
Kelly,~K.; Coronado,~E.; Zhao,~L.; Schatz,~G. The Optical Properties of Metal
  Nanoparticles: the Influence of Size, Shape, and Dielectric Environment.
  \emph{J.~Phys.~Chem.~B} \textbf{2003}, \emph{107}, 668--677\relax
\mciteBstWouldAddEndPuncttrue
\mciteSetBstMidEndSepPunct{\mcitedefaultmidpunct}
{\mcitedefaultendpunct}{\mcitedefaultseppunct}\relax
\EndOfBibitem
\bibitem[Barnes et~al.(2003)Barnes, Dereux, and Ebbesen]{barn+03nat}
Barnes,~W.; Dereux,~A.; Ebbesen,~T. Surface Plasmon Subwavelength Optics.
  \emph{Nature} \textbf{2003}, \emph{424}, 824--830\relax
\mciteBstWouldAddEndPuncttrue
\mciteSetBstMidEndSepPunct{\mcitedefaultmidpunct}
{\mcitedefaultendpunct}{\mcitedefaultseppunct}\relax
\EndOfBibitem
\bibitem[Lal et~al.(2007)Lal, Link, and Halas]{lal+07natpho}
Lal,~S.; Link,~S.; Halas,~N. Nano-Optics from Sensing to Waveguiding.
  \emph{Nat. Photonics} \textbf{2007}, \emph{1}, 641--648\relax
\mciteBstWouldAddEndPuncttrue
\mciteSetBstMidEndSepPunct{\mcitedefaultmidpunct}
{\mcitedefaultendpunct}{\mcitedefaultseppunct}\relax
\EndOfBibitem
\bibitem[Konstantatos and Sargent(2010)]{kons-sarg10natn}
Konstantatos,~G.; Sargent,~E. Nanostructured Materials for Photon Detection.
  \emph{Nature nanotechnology} \textbf{2010}, \emph{5}, 391--400\relax
\mciteBstWouldAddEndPuncttrue
\mciteSetBstMidEndSepPunct{\mcitedefaultmidpunct}
{\mcitedefaultendpunct}{\mcitedefaultseppunct}\relax
\EndOfBibitem
\bibitem[West et~al.(2010)West, Ishii, Naik, Emani, Shalev, and
  Boltasseva]{west+10lpr}
West,~P.; Ishii,~S.; Naik,~G.; Emani,~N.; Shalev,~V.; Boltasseva,~A. Searching
  for Better Plasmonic Materials. \emph{Laser~Photonics~Rev.~} \textbf{2010},
  \emph{4}, 795--808\relax
\mciteBstWouldAddEndPuncttrue
\mciteSetBstMidEndSepPunct{\mcitedefaultmidpunct}
{\mcitedefaultendpunct}{\mcitedefaultseppunct}\relax
\EndOfBibitem
\bibitem[Castro~Neto et~al.(2009)Castro~Neto, Guinea, Peres, Novoselov, and
  Geim]{cast+09rmp}
Castro~Neto,~A.~H.; Guinea,~F.; Peres,~N. M.~R.; Novoselov,~K.~S.; Geim,~A.~K.
  The electronic properties of graphene. \emph{Rev.~Mod.~Phys.~} \textbf{2009},
  \emph{81}, 109--162\relax
\mciteBstWouldAddEndPuncttrue
\mciteSetBstMidEndSepPunct{\mcitedefaultmidpunct}
{\mcitedefaultendpunct}{\mcitedefaultseppunct}\relax
\EndOfBibitem
\bibitem[Benisty(2009)]{beni09prb}
Benisty,~H. Graphene Nanoribbons: Photonic Crystal Waveguide Analogy and
  Minigap Stripes. \emph{Phys. Rev. B} \textbf{2009}, \emph{79}, 155409\relax
\mciteBstWouldAddEndPuncttrue
\mciteSetBstMidEndSepPunct{\mcitedefaultmidpunct}
{\mcitedefaultendpunct}{\mcitedefaultseppunct}\relax
\EndOfBibitem
\bibitem[Avouris(2010)]{avou10nl}
Avouris,~P. Graphene: Electronic and Photonic Properties and Devices.
  \emph{Nano~Lett.~} \textbf{2010}, \emph{10}, 4285--4294\relax
\mciteBstWouldAddEndPuncttrue
\mciteSetBstMidEndSepPunct{\mcitedefaultmidpunct}
{\mcitedefaultendpunct}{\mcitedefaultseppunct}\relax
\EndOfBibitem
\bibitem[Bonaccorso et~al.(2010)Bonaccorso, Sun, Hasan, and
  Ferrari]{bona+10natpho}
Bonaccorso,~F.; Sun,~Z.; Hasan,~T.; Ferrari,~A. Graphene Photonics and
  Optoelectronics. \emph{Nat. Photonics} \textbf{2010}, \emph{4},
  611--622\relax
\mciteBstWouldAddEndPuncttrue
\mciteSetBstMidEndSepPunct{\mcitedefaultmidpunct}
{\mcitedefaultendpunct}{\mcitedefaultseppunct}\relax
\EndOfBibitem
\bibitem[Hwang and Sarma(2007)]{hwan-sarm07prb}
Hwang,~E.; Sarma,~S. Dielectric Function, Screening, and Plasmons in
  Two-Dimensional Graphene. \emph{Phys.~Rev.~B} \textbf{2007}, \emph{75},
  205418\relax
\mciteBstWouldAddEndPuncttrue
\mciteSetBstMidEndSepPunct{\mcitedefaultmidpunct}
{\mcitedefaultendpunct}{\mcitedefaultseppunct}\relax
\EndOfBibitem
\bibitem[Eberlein et~al.(2008)Eberlein, Bangert, Nair, Jones, Gass, Bleloch,
  Novoselov, Geim, and Briddon]{erbe+08prb}
Eberlein,~T.; Bangert,~U.; Nair,~R.~R.; Jones,~R.; Gass,~M.; Bleloch,~A.~L.;
  Novoselov,~K.~S.; Geim,~A.; Briddon,~P.~R. Plasmon Spectroscopy of
  Free-Standing Graphene Films. \emph{Phys. Rev. B} \textbf{2008}, \emph{77},
  233406\relax
\mciteBstWouldAddEndPuncttrue
\mciteSetBstMidEndSepPunct{\mcitedefaultmidpunct}
{\mcitedefaultendpunct}{\mcitedefaultseppunct}\relax
\EndOfBibitem
\bibitem[Kramberger et~al.(2008)Kramberger, Hambach, Giorgetti, R\"ummeli,
  Knupfer, Fink, B\"uchner, Reining, Einarsson, Maruyama, Sottile, Hannewald,
  Olevano, Marinopoulos, and Pichler]{kram+08prl}
Kramberger,~C.; Hambach,~R.; Giorgetti,~C.; R\"ummeli,~M.~H.; Knupfer,~M.;
  Fink,~J.; B\"uchner,~B.; Reining,~L.; Einarsson,~E.; Maruyama,~S. \emph{et al.}
  Linear Plasmon Dispersion in Single-Wall Carbon Nanotubes and the Collective
  Excitation Spectrum of Graphene. \emph{Phys. Rev. Lett.} \textbf{2008},
  \emph{100}, 196803\relax
\mciteBstWouldAddEndPuncttrue
\mciteSetBstMidEndSepPunct{\mcitedefaultmidpunct}
{\mcitedefaultendpunct}{\mcitedefaultseppunct}\relax
\EndOfBibitem
\bibitem[Gass et~al.(2008)Gass, Bangert, Bleloch, Wang, Nair, and
  Geim]{gass+08natn}
Gass,~M.; Bangert,~U.; Bleloch,~A.; Wang,~P.; Nair,~R.; Geim,~A. Free-Standing
  Graphene at Atomic Resolution. \emph{Nature Nanotech.} \textbf{2008},
  \emph{3}, 676--681\relax
\mciteBstWouldAddEndPuncttrue
\mciteSetBstMidEndSepPunct{\mcitedefaultmidpunct}
{\mcitedefaultendpunct}{\mcitedefaultseppunct}\relax
\EndOfBibitem
\bibitem[Jablan et~al.(2009)Jablan, Buljan, and Solja\ifmmode \check{c}\else
  \v{c}\fi{}i\ifmmode~\acute{c}\else \'{c}\fi{}]{jabl+09prb}
Jablan,~M.; Buljan,~H.; Solja\ifmmode \check{c}\else
  \v{c}\fi{}i\ifmmode~\acute{c}\else \'{c}\fi{},~M. Plasmonics in Graphene at
  Infrared Frequencies. \emph{Phys. Rev. B} \textbf{2009}, \emph{80},
  245435\relax
\mciteBstWouldAddEndPuncttrue
\mciteSetBstMidEndSepPunct{\mcitedefaultmidpunct}
{\mcitedefaultendpunct}{\mcitedefaultseppunct}\relax
\EndOfBibitem
\bibitem[Mishchenko et~al.(2010)Mishchenko, Shytov, and Silvestrov]{mish+10prl}
Mishchenko,~E.~G.; Shytov,~A.~V.; Silvestrov,~P.~G. Guided Plasmons in Graphene
  $p-n$ Junctions. \emph{Phys. Rev. Lett.} \textbf{2010}, \emph{104},
  156806\relax
\mciteBstWouldAddEndPuncttrue
\mciteSetBstMidEndSepPunct{\mcitedefaultmidpunct}
{\mcitedefaultendpunct}{\mcitedefaultseppunct}\relax
\EndOfBibitem
\bibitem[Koppens et~al.(2011)Koppens, Chang, and Garcia~de Abajo]{kopp+11nl}
Koppens,~F. H.~L.; Chang,~D.~E.; Garc\'ia~de Abajo,~F.~J. Graphene Plasmonics: A
  Platform for Strong Light--Matter Interactions. \emph{Nano~Lett.~}
  \textbf{2011}, \emph{11}, 3370--3377\relax
\mciteBstWouldAddEndPuncttrue
\mciteSetBstMidEndSepPunct{\mcitedefaultmidpunct}
{\mcitedefaultendpunct}{\mcitedefaultseppunct}\relax
\EndOfBibitem
\bibitem[Thongrattanasiri et~al.()Thongrattanasiri, Manjavacas, and Garci­a~de
  Abajo]{thon+12nano}
Thongrattanasiri,~S.; Manjavacas,~A.; Garc\'ia~de Abajo,~F.~J. Quantum
  Finite-Size Effects in Graphene Plasmons. \emph{ACS Nano}   
  \textbf{2012}, \emph{6}, 1766--1775\relax
\mciteBstWouldAddEndPuncttrue
\mciteSetBstMidEndSepPunct{\mcitedefaultmidpunct}
{\mcitedefaultendpunct}{\mcitedefaultseppunct}\relax
\EndOfBibitem
\bibitem[Nikitin et~al.(2011)Nikitin, Guinea, Garc\'ia-Vidal, and
  Mart\'in-Moreno]{niki+11prb}
Nikitin,~A.~Y.; Guinea,~F.; Garc\'ia-Vidal,~F.~J.; Mart\'in-Moreno,~L. Edge and
  Waveguide Terahertz Surface Plasmon Modes in Graphene Microribbons.
  \emph{Phys.~Rev.~B} \textbf{2011}, \emph{84}, 161407\relax
\mciteBstWouldAddEndPuncttrue
\mciteSetBstMidEndSepPunct{\mcitedefaultmidpunct}
{\mcitedefaultendpunct}{\mcitedefaultseppunct}\relax
\EndOfBibitem
\bibitem[Vakil and Engheta(2011)]{vaki-engh11sci}
Vakil,~A.; Engheta,~N. Transformation Optics Using Graphene. \emph{Science}
  \textbf{2011}, \emph{332}, 1291--1294\relax
\mciteBstWouldAddEndPuncttrue
\mciteSetBstMidEndSepPunct{\mcitedefaultmidpunct}
{\mcitedefaultendpunct}{\mcitedefaultseppunct}\relax
\EndOfBibitem
\bibitem[Wang et~al.(2011)Wang, Apell, and Kinaret]{wang+11prb}
Wang,~W.; Apell,~P.; Kinaret,~J. Edge Plasmons in Graphene Nanostructures.
  \emph{Phys.~Rev.~B} \textbf{2011}, \emph{84}, 085423\relax
\mciteBstWouldAddEndPuncttrue
\mciteSetBstMidEndSepPunct{\mcitedefaultmidpunct}
{\mcitedefaultendpunct}{\mcitedefaultseppunct}\relax
\EndOfBibitem
\bibitem[Ju et~al.(2011)Ju, Geng, Horng, Girit, Martin, Hao, Bechtel, Liang,
  Zettl, Shen,et~al.]{ju+11natn}
Ju,~L.; Geng,~B.; Horng,~J.; Girit,~C.; Martin,~M.; Hao,~Z.; Bechtel,~H.;
  Liang,~X.; Zettl,~A.; Shen,~Y. \emph{et~al.} Graphene Plasmonics for Tunable Terahertz
  Metamaterials. \emph{Nature Nanotech.} \textbf{2011}, \emph{6},
  630--634\relax
\mciteBstWouldAddEndPuncttrue
\mciteSetBstMidEndSepPunct{\mcitedefaultmidpunct}
{\mcitedefaultendpunct}{\mcitedefaultseppunct}\relax
\EndOfBibitem
\bibitem[Christensen et~al.(2012)Christensen, Manjavacas, Thongrattanasiri,
  Koppens, and Garc\'ia~de Abajo]{chri+12nano}
Christensen,~J.; Manjavacas,~A.; Thongrattanasiri,~S.; Koppens,~F. H.~L.;
  Garc\'ia~de Abajo,~F.~J. Graphene Plasmon Waveguiding and Hybridization in
  Individual and Paired Nanoribbons. \emph{ACS~Nano} \textbf{2012}, \emph{6},
  431--440\relax
\mciteBstWouldAddEndPuncttrue
\mciteSetBstMidEndSepPunct{\mcitedefaultmidpunct}
{\mcitedefaultendpunct}{\mcitedefaultseppunct}\relax
\EndOfBibitem
\bibitem[Wu et~al.(2007)Wu, Pisula, and Muellen]{wu+07cr}
Wu,~J.; Pisula,~W.; Muellen,~K. Graphenes as Potential Material for
  Electronics. \emph{Chem.~Rev.~} \textbf{2007}, \emph{107}, 718--747\relax
\mciteBstWouldAddEndPuncttrue
\mciteSetBstMidEndSepPunct{\mcitedefaultmidpunct}
{\mcitedefaultendpunct}{\mcitedefaultseppunct}\relax
\EndOfBibitem
\bibitem[Yang et~al.(2008)Yang, Dou, Rouhanipour, Zhi, Rader, and
  Muellen]{yang+08jacs}
Yang,~X.; Dou,~X.; Rouhanipour,~A.; Zhi,~L.; Rader,~H.~J.; Muellen,~K.
  Two-Dimensional Graphene Nanoribbons. \emph{J.~Am.~Chem.~Soc.~}
  \textbf{2008}, \emph{130}, 4216\relax
\mciteBstWouldAddEndPuncttrue
\mciteSetBstMidEndSepPunct{\mcitedefaultmidpunct}
{\mcitedefaultendpunct}{\mcitedefaultseppunct}\relax
\EndOfBibitem
\bibitem[Cai et~al.(2010)Cai, Ruffieux, Jaafar, Bieri, Braun, Blankenburg,
  Muoth, Saleh, Feng, Muellen, and Fasel]{cai+10nat}
Cai,~J.; Ruffieux,~P.; Jaafar,~R.; Bieri,~M.; Braun,~T.; Blankenburg,~S.;
  Muoth,~M.; Seitsonen,~A.~P.; Saleh,~M.; Feng,~X. \emph{et al.} Atomically Precise Bottom-Up Fabrication of Graphene Nanoribbons.
  \emph{Nature~(London)~} \textbf{2010}, \emph{466}, 470--473\relax
\mciteBstWouldAddEndPuncttrue
\mciteSetBstMidEndSepPunct{\mcitedefaultmidpunct}
{\mcitedefaultendpunct}{\mcitedefaultseppunct}\relax
\EndOfBibitem
\bibitem[Palma and Samor{\'i}(2011)]{palm-samo11natc}
Palma,~C.; Samor{\'i},~P. Blueprinting Macromolecular Electronics. \emph{Nature
  Chem.} \textbf{2011}, \emph{3}, 431--436\relax
\mciteBstWouldAddEndPuncttrue
\mciteSetBstMidEndSepPunct{\mcitedefaultmidpunct}
{\mcitedefaultendpunct}{\mcitedefaultseppunct}\relax
\EndOfBibitem
\bibitem[Nakada et~al.(1996)Nakada, Fujita, Dresselhaus, and
  Dresselhaus]{naka+96prb}
Nakada,~K.; Fujita,~M.; Dresselhaus,~G.; Dresselhaus,~M.~S. Edge State in
  Graphene Ribbons: Nanometer Size Effect and Edge Shape Dependence.
  \emph{Phys.~Rev.~B} \textbf{1996}, \emph{54}, 17954--17961\relax
\mciteBstWouldAddEndPuncttrue
\mciteSetBstMidEndSepPunct{\mcitedefaultmidpunct}
{\mcitedefaultendpunct}{\mcitedefaultseppunct}\relax
\EndOfBibitem
\bibitem[Ridley and Zerner(1973)]{ridl-zern73tca}
Ridley,~J.; Zerner,~M. An Intermediate Neglect of Differential Overlap
  Technique for Spectroscopy: Pyrrole and the Azines. \emph{Theor.~Chem.~Acta}
  \textbf{1973}, \emph{32}, 111--134\relax
\mciteBstWouldAddEndPuncttrue
\mciteSetBstMidEndSepPunct{\mcitedefaultmidpunct}
{\mcitedefaultendpunct}{\mcitedefaultseppunct}\relax
\EndOfBibitem
\bibitem[Caldas et~al.(2001)Caldas, Pettenati, Goldoni, and
  Molinari]{cald+01apl}
Caldas,~M.~J.; Pettenati,~E.; Goldoni,~G.; Molinari,~E. Tailoring of Light
  Emission Properties of Functionalized Oligothiophenes.
  \emph{Appl.~Phys.~Lett.~} \textbf{2001}, \emph{79}, 2505--2507\relax
\mciteBstWouldAddEndPuncttrue
\mciteSetBstMidEndSepPunct{\mcitedefaultmidpunct}
{\mcitedefaultendpunct}{\mcitedefaultseppunct}\relax
\EndOfBibitem
\bibitem[Wang et~al.(2004)Wang, Tomovi\'{c}, Kastler, Pretsch, Negri,
  Enkelmann, and Muellen]{wang+04jacs}
Wang,~Z.; Tomovi\'{c},~Z.; Kastler,~M.; Pretsch,~R.; Negri,~F.; Enkelmann,~V.;
  Muellen,~K. Graphitic Molecules with Partial "Zig/Zag" Periphery.
  \emph{J.~Am.~Chem.~Soc.~} \textbf{2004}, \emph{126}, 7794--7795\relax
\mciteBstWouldAddEndPuncttrue
\mciteSetBstMidEndSepPunct{\mcitedefaultmidpunct}
{\mcitedefaultendpunct}{\mcitedefaultseppunct}\relax
\EndOfBibitem
\bibitem[Cocchi et~al.(2011)Cocchi, Prezzi, Ruini, Caldas, and
  Molinari]{cocc+11jpcl}
Cocchi,~C.; Prezzi,~D.; Ruini,~A.; Caldas,~M.~J.; Molinari,~E. Optical
  Properties and Charge-Transfer Excitations in Edge-Functionalized
  All-Graphene Nanojunctions. \emph{J.~Phys.~Chem.~Lett.~} \textbf{2011},
  \emph{2}, 1315--1319\relax
\mciteBstWouldAddEndPuncttrue
\mciteSetBstMidEndSepPunct{\mcitedefaultmidpunct}
{\mcitedefaultendpunct}{\mcitedefaultseppunct}\relax
\EndOfBibitem
\bibitem[Not()]{Note-1}
 The reliability of the ZINDO model for the target systems was further
  demonstrated by computing time-dependent density-functional theory
  (TDDFT-b3lyp) excitation energies and oscillator strengths for a selected
  flake of length $\sim$ 24 \AA{} (GAMESS package with 3-21G basis set:
  \url{http://www.msg.ameslab.gov/gamess/})\relax
\mciteBstWouldAddEndPuncttrue
\mciteSetBstMidEndSepPunct{\mcitedefaultmidpunct}
{\mcitedefaultendpunct}{\mcitedefaultseppunct}\relax
\EndOfBibitem
\bibitem[Dewar et~al.(1985)Dewar, Zoebish, Healy, and Stewart]{dewa+85jacs}
Dewar,~M. J.~S.; Zoebish,~E.~G.; Healy,~E.~F.; Stewart,~J. J.~P. A New General
  Purpose Quantum Mechanical Molecular Model. \emph{J.~Am.~Chem.~Soc.~}
  \textbf{1985}, \emph{107}, 3902--3909\relax
\mciteBstWouldAddEndPuncttrue
\mciteSetBstMidEndSepPunct{\mcitedefaultmidpunct}
{\mcitedefaultendpunct}{\mcitedefaultseppunct}\relax
\EndOfBibitem
\bibitem[Cocchi et~al.(2011)Cocchi, Ruini, Prezzi, Caldas, and
  Molinari]{cocc+11jpcc}
Cocchi,~C.; Ruini,~A.; Prezzi,~D.; Caldas,~M.~J.; Molinari,~E. Designing
  All-Graphene Nanojunctions by Covalent Functionalization.
  \emph{J.~Phys.~Chem.~C} \textbf{2011}, \emph{115}, 2969--2973\relax
\mciteBstWouldAddEndPuncttrue
\mciteSetBstMidEndSepPunct{\mcitedefaultmidpunct}
{\mcitedefaultendpunct}{\mcitedefaultseppunct}\relax
\EndOfBibitem
\bibitem[Shemella et~al.(2007)Shemella, Zhang, Mailman, Ajayan, and
  Nayak]{shem+07apl}
Shemella,~P.; Zhang,~Y.; Mailman,~M.; Ajayan,~P.; Nayak,~S. Energy Gaps in
  Zero-Dimensional Graphene Nanoribbons. \emph{Appl.~Phys.~Lett.~}
  \textbf{2007}, \emph{91}, 042101\relax
\mciteBstWouldAddEndPuncttrue
\mciteSetBstMidEndSepPunct{\mcitedefaultmidpunct}
{\mcitedefaultendpunct}{\mcitedefaultseppunct}\relax
\EndOfBibitem
\bibitem[Hod et~al.(2008)Hod, Barone, and Scuseria]{hod+08prb}
Hod,~O.; Barone,~V.; Scuseria,~G.~E. Half-Metallic Graphene Nanodots: A
  Comprehensive First-Principles Theoretical Study. \emph{Phys.~Rev.~B}
  \textbf{2008}, \emph{77}, 035411\relax
\mciteBstWouldAddEndPuncttrue
\mciteSetBstMidEndSepPunct{\mcitedefaultmidpunct}
{\mcitedefaultendpunct}{\mcitedefaultseppunct}\relax
\EndOfBibitem
\bibitem[Prezzi et~al.(2008)Prezzi, Varsano, Ruini, Marini, and
  Molinari]{prez+08prb}
Prezzi,~D.; Varsano,~D.; Ruini,~A.; Marini,~A.; Molinari,~E. Optical Properties
  of Graphene Nanoribbons: The Role of Many-Body Effects. \emph{Phys.~Rev.~B}
  \textbf{2008}, \emph{77}, 041404(R)\relax
\mciteBstWouldAddEndPuncttrue
\mciteSetBstMidEndSepPunct{\mcitedefaultmidpunct}
{\mcitedefaultendpunct}{\mcitedefaultseppunct}\relax
\EndOfBibitem
\bibitem[Not()]{Note-2}
 The approximations included within the adopted ZINDO/S approach, specifically
  in the neglect of two center integrals, have to be taken into account in the
  calculation of $\rho^{0p} (\rv)$.\relax
\mciteBstWouldAddEndPunctfalse
\mciteSetBstMidEndSepPunct{\mcitedefaultmidpunct}
{}{\mcitedefaultseppunct}\relax
\EndOfBibitem
\bibitem[Yan et~al.(2007)Yan, Yuan, and Gao]{yan+07prl}
Yan,~J.; Yuan,~Z.; Gao,~S. End and Central Plasmon Resonances in Linear Atomic
  Chains. \emph{Phys. Rev. Lett.} \textbf{2007}, \emph{98}, 216602\relax
\mciteBstWouldAddEndPuncttrue
\mciteSetBstMidEndSepPunct{\mcitedefaultmidpunct}
{\mcitedefaultendpunct}{\mcitedefaultseppunct}\relax
\EndOfBibitem
\bibitem[Yan and Gao(2008)]{yan-gao08prb}
Yan,~J.; Gao,~S. Plasmon Resonances in Linear Atomic Chains: Free-Electron
  Behavior and Anisotropic Screening of $d$ Electrons. \emph{Phys. Rev. B}
  \textbf{2008}, \emph{78}, 235413\relax
\mciteBstWouldAddEndPuncttrue
\mciteSetBstMidEndSepPunct{\mcitedefaultmidpunct}
{\mcitedefaultendpunct}{\mcitedefaultseppunct}\relax
\EndOfBibitem
\bibitem[Lian et~al.(2009)Lian, Sa{\l}ek, Jin, and Ding]{lian+09jcp}
Lian,~K.; Sa{\l}ek,~P.; Jin,~M.; Ding,~D. Density-Functional Studies of
  Plasmons in Small Metal Clusters. \emph{J.~Chem.~Phys.~} \textbf{2009},
  \emph{130}, 174701\relax
\mciteBstWouldAddEndPuncttrue
\mciteSetBstMidEndSepPunct{\mcitedefaultmidpunct}
{\mcitedefaultendpunct}{\mcitedefaultseppunct}\relax
\EndOfBibitem
\bibitem[Yasuike et~al.(2011)Yasuike, Nobusada, and Hayashi]{yasu+11pra}
Yasuike,~T.; Nobusada,~K.; Hayashi,~M. Collectivity of Plasmonic Excitations in
  Small Sodium Clusters with Ring and Linear Structures. \emph{Phys. Rev. A}
  \textbf{2011}, \emph{83}, 013201\relax
\mciteBstWouldAddEndPuncttrue
\mciteSetBstMidEndSepPunct{\mcitedefaultmidpunct}
{\mcitedefaultendpunct}{\mcitedefaultseppunct}\relax
\EndOfBibitem
\bibitem[Scholes and Rumbles(2006)]{scho06natm}
Scholes,~G.; Rumbles,~G. Excitons in Nanoscale Systems. \emph{Nature Mat.}
  \textbf{2006}, \emph{5}, 683--696\relax
\mciteBstWouldAddEndPuncttrue
\mciteSetBstMidEndSepPunct{\mcitedefaultmidpunct}
{\mcitedefaultendpunct}{\mcitedefaultseppunct}\relax
\EndOfBibitem
\bibitem[Matsuda et~al.(2008)Matsuda, Inoue, Murakami, Maruyama, and
  Kanemitsu]{mats+08prb}
Matsuda,~K.; Inoue,~T.; Murakami,~Y.; Maruyama,~S.; Kanemitsu,~Y. Exciton
  Dephasing and Multiexciton Recombinations in a Single Carbon Nanotube.
  \emph{Phys.~Rev.~B} \textbf{2008}, \emph{77}, 033406\relax
\mciteBstWouldAddEndPuncttrue
\mciteSetBstMidEndSepPunct{\mcitedefaultmidpunct}
{\mcitedefaultendpunct}{\mcitedefaultseppunct}\relax
\EndOfBibitem
\bibitem[Son et~al.(2006)Son, Cohen, and Louie]{son+06prl}
Son,~Y.-W.; Cohen,~M.~L.; Louie,~S.~G. Energy Gaps in Graphene Nanoribbons.
  \emph{Phys.~Rev.~Lett.~} \textbf{2006}, \emph{97}, 216803\relax
\mciteBstWouldAddEndPuncttrue
\mciteSetBstMidEndSepPunct{\mcitedefaultmidpunct}
{\mcitedefaultendpunct}{\mcitedefaultseppunct}\relax
\EndOfBibitem
\bibitem[Barone et~al.(2006)Barone, Hod, and Scuseria]{baro+06nl}
Barone,~V.; Hod,~O.; Scuseria,~G.~E. Electronic Structure and Stability of
  Semiconducting Graphene Nanoribbons. \emph{Nano~Lett.~} \textbf{2006},
  \emph{6}, 2748--2754\relax
\mciteBstWouldAddEndPuncttrue
\mciteSetBstMidEndSepPunct{\mcitedefaultmidpunct}
{\mcitedefaultendpunct}{\mcitedefaultseppunct}\relax
\EndOfBibitem
\bibitem[Yang et~al.(2007)Yang, Cohen, and Louie]{yang+07nl}
Yang,~L.; Cohen,~M.; Louie,~S. Excitonic Effects in the Optical Spectra of
  Graphene Nanoribbons. \emph{Nano~Lett.~} \textbf{2007}, \emph{7},
  3112--3115\relax
\mciteBstWouldAddEndPuncttrue
\mciteSetBstMidEndSepPunct{\mcitedefaultmidpunct}
{\mcitedefaultendpunct}{\mcitedefaultseppunct}\relax
\EndOfBibitem
\bibitem[Prezzi et~al.(2011)Prezzi, Varsano, Ruini, and Molinari]{prez+11prb}
Prezzi,~D.; Varsano,~D.; Ruini,~A.; Molinari,~E. Quantum Dot States and Optical
  Excitations of Edge-Modulated Graphene Nanoribbons. \emph{Phys.~Rev.~B}
  \textbf{2011}, \emph{84}, 041401\relax
\mciteBstWouldAddEndPuncttrue
\mciteSetBstMidEndSepPunct{\mcitedefaultmidpunct}
{\mcitedefaultendpunct}{\mcitedefaultseppunct}\relax
\EndOfBibitem
\end{mcitethebibliography}
\end{document}